\documentclass[aps,pre,reprint,groupedaddress,longbibliography]{revtex4-2}

\usepackage{graphicx}
\usepackage{dcolumn}
\usepackage{amsmath}
\usepackage{bm}
\usepackage{physics}
\usepackage[usenames,dvipsnames]{xcolor}
\usepackage[normalem]{ulem}

\newcommand{\add}[1]{{#1}} 
\newcommand{\stkout}[1]{} 

\newcommand{\Eddag}{E^\ddag}

\begin{document}

\title{Multiparameter optimal control of $\mathrm{F_1}$-ATPase}
\author{W.\ Callum Wareham}
\email{callum\_wareham@sfu.ca}
\author{David A.\ Sivak}
\email{dsivak@sfu.ca}
\affiliation{Department of Physics, Simon Fraser University, Burnaby, British Columbia V5A 1S6, Canada} 
\date{\today}

\begin{abstract}
Biological molecular machines convert free energy between different forms in cells, often at high efficiency. Optimal control theory provides a framework to elucidate design principles governing energetically efficient driving. Here, we use linear-response theory to design efficient protocols exercising dynamic control of trap center and stiffness in a model of driven $\mathrm{F_1}$-ATPase. We find that the key design principles of an efficient protocol can be satisfied either by dynamic control of both parameters or by dynamic control of a single parameter and a good static choice for the second. These results illustrate that accessing a new degree of dynamic control provides varying performance improvements in different systems.
\end{abstract}

\maketitle

\section{Introduction}

Biological cells must constantly convert energy between different forms to function and survive~\cite{leightonFlowEnergyInformation2024}. Usually, energy conversion is tasked to nanoscale protein complexes known as molecular machines~\cite{brownTheoryNonequilibriumFree2020}. Various examples include the proteins responsible for light harvesting during photosynthesis~\cite{vinyardPhotosystemIIReaction2013a}, for converting the chemical energy in food to a common energy currency~\cite{okunoRotationStructureFoF1ATP2011a}, and for directed motion within or by cells~\cite{sweeneyMotorProteins2018,nirodyBiophysicistsGuideBacterial2017,guoBacterialFlagellarMotor2022}. Evolutionarily, there is a selective pressure toward energetic efficiency for these machines.

Indeed, the rotary motor $\mathrm{F_o}$-$\mathrm{F_1}$ adenosine triphosphate (ATP) synthase, which is the last step in cellular respiration in bacterial plasma membranes and eukaryotic mitochondria, converts its input $\rm H^+$ chemical-potential difference into ATP (a universal cellular energy currency) at high efficiencies of 70--90\%~\cite{silversteinExplorationHowThermodynamic2014}. This efficiency is achieved despite the strong stochastic fluctuations the machine faces, and the fact that it has negligible inertia to carry it through any steps of its cycle~\cite{brownTheoryNonequilibriumFree2020}.

Taking inspiration from the natural world, synthetic molecular machines have been developed \textit{in vitro}~\cite{zhangArtificialMolecularMachines2023}. These machines are already extraordinary successes of nano-engineering, but their efficiencies remain orders of magnitude lower than their biological counterparts~\cite{amanoInsightsInformationThermodynamics2022}. In the tradition of biomimicry, it is natural to look to the design principles governing the efficiency of biological molecular machines in the face of their distinctive challenges.

A popular line of inquiry into the efficient dynamics of mesoscopic stochastic systems has been through optimal control theory~\cite{aurellOptimalProtocolsOptimal2011,blaberOptimalControlStochastic2023}. Often, molecular machines are modeled as one protein unit (the controller) that uses an upstream energy reservoir to manipulate a few \textit{control parameters}, which change the potential-energy landscape that a downstream unit experiences. Given the goals and structure of the downstream component, the optimal control problem asks what schedule (or \textit{control protocol}) the control parameters should follow to drive the downstream unit while avoiding \textit{excess work} beyond the equilibrium change in free energy. Even for relatively simple downstream potentials, general solutions to the optimal control problem are difficult to find, so approximations have been developed in various limits~\cite{blaberOptimalControlStochastic2023}.

The near-equilibrium, linear-response limit yields a formalism known as thermodynamic geometry~\cite{sivakThermodynamicMetricsOptimal2012,zulkowskiGeometryThermodynamicControl2012}, which has proven fruitful in a range of models of stochastic systems\add{, even surprisingly far from equilibrium}~\cite{rotskoffOptimalControlNonequilibrium2015,guptaOptimalControlF1ATPase2022,blaberEfficientTwodimensionalControl2022,louwerseMultidimensionalMinimumworkControl2022}. Recently, Ref.~\cite{guptaOptimalControlF1ATPase2022} applied this formalism to a model of an \textit{in vitro} control experiment in which the downstream element $\mathrm{F_1}$ is connected to a micron-sized bead and rotated by a magnetic trap. Calculating designed protocols with this framework reduced the excess work in driving ATP synthesis by up to $\approx$2.5$\times$ compared to the \emph{naive} protocol (that with constant angular velocity) in both theory and simulations. 

Like many other investigations, however, Ref.~\cite{guptaOptimalControlF1ATPase2022} dynamically varied a single control parameter (the rotational angle of the magnetic trap). Meanwhile, multiparameter control protocols have been found to yield large savings above and beyond their single-parameter counterparts, notably in the similar case of a double-well potential~\cite{blaberEfficientTwodimensionalControl2022,blaberOptimalControlStochastic2023}. In Ref.~\cite{blaberEfficientTwodimensionalControl2022}, protocols using dynamic control of the center and stiffness of a quadratic trap reduced excess work done by the trap by $\approx$3.5$\times$ compared to those designed to control only the trap center. In light of these results, it is natural to ask if the large additional savings that trap-stiffness control provides are a general feature, including in models that more faithfully reproduce the real-world dynamics of molecular machines.

Here, we extend the results of Ref.~\cite{guptaOptimalControlF1ATPase2022} to design control protocols that dynamically vary both the trap-center and trap-stiffness control parameters. To our knowledge, this is the first direct application of multiparameter optimal control theory to a widely used model for a biological molecular machine~\cite{kawaguchiNonequilibriumDissipationfreeTransport2014}. We find that control protocols that apply a static, strong trap and dynamically vary only the trap center can achieve nearly the same performance as protocols making dynamic use of both coordinates. The strong-trap, single-parameter protocols satisfy the key aspects (the ``stiff'' modes~\cite{machtaParameterSpaceCompression2013}) for minimizing the excess work, so the additional control exercised by the second parameter provides little additional savings. For all these protocols, we also find that the true excess work computed with Langevin simulations compares well with that predicted by the linear-response framework. Our results illustrate that different aspects of control can differ dramatically in their utility, and that adding additional degrees of dynamic control does not guarantee a large improvement in efficiency.

\section{Theory}

We model an $\mathrm{F_1}$ control experiment with the same scheme used in Ref.~\cite{guptaOptimalControlF1ATPase2022} that builds on the widely used $\mathrm{F_1}$ model from \cite{kawaguchiNonequilibriumDissipationfreeTransport2014}, which we summarize here for completeness. $\mathrm{F_1}$ has threefold symmetry, and its rotor crankshaft takes discrete steps of $\xi = 1/3$ rotations when operating in either ATP synthesis or hydrolysis mode. Each of these steps is composed of two substeps, \stkout{separated by a distance $\ell= 1/9$ rot} \add{of 
sizes
$\ell = 1/9$ and $2/9$ rot~\cite{yasudaF1ATPaseHighlyEfficient1998,yasudaResolutionDistinctRotational2001,hirono-haraPauseRotationF1ATPase2001,shimabukuroCatalysisRotationF12003, uenoATPdrivenStepwiseRotation2005, okunoRotationStructureFoF1ATP2011a}}. The model, therefore, includes six distinct rotational substates, each coupled to the bead position via a quadratic potential with the same shape. In the model, two consecutive substates are coarse-grained into a single state\add{~\cite[Fig.~S1]{kawaguchiNonequilibriumDissipationfreeTransport2014}}, yielding a potential of mean force as a function of angle:
\begin{equation}
    \beta U_n(\theta) = 
    -\ln\left[e^{-\beta k(\theta+\ell - n\xi)^2/2-\beta \Delta\tilde{\mu}} + e^{-\beta k(\theta-n\xi)^2/2}\right]\stkout{.}, \label{eq:state_coarsegrained}
\end{equation}
\add{where $\beta \equiv 1/k_\mathrm{B} T$.}
The integer $n$ indexes $\mathrm{F_1}$'s rotational state, $\theta$ is the bead angle, $k$$\approx$790$~k_{\rm B} T/\mathrm{rot}^2$ is the stiffness of a single ATP-synthase rotational substate, and $\Delta\tilde{\mu} = 5.2~k_{\rm B} T$ is the free-energy difference between the two consecutive substates.

To obey the local detailed-balance conditions~\cite{guptaOptimalControlF1ATPase2022,kawaguchiNonequilibriumDissipationfreeTransport2014}, the rotational state \add{(and thus the currently active potential-energy surface $U_n$)} switches at hydrolysis rate $R^+_n$ and synthesis rate $R^-_n$, which are related by the ratio
\begin{equation}
    \frac{R_n^+(\theta)}{R_{n+1}^-(\theta)} = e^{\beta[U_n(\theta) - U_{n+1}(\theta) + \Delta \mu_{\mathrm{ATP}}]},
\end{equation}
where $\Delta \mu_{\mathrm{ATP}} = 18~k_{\rm B} T$ is the chemical-potential change for the synthesis of one ATP. The rate constant $\Gamma = R_n^+(\theta)$ increases along with [ATP] and [ADP]. For low chemical concentrations (and hence slow chemical dynamics compared to mechanical motion), $\Gamma$ is small, and the rotor stalk mechanically fluctuates while waiting for a chemical reaction to occur; meanwhile, it experiences the potential from one rotational state (i.e., one value of $n$). Near the intersection of the potentials contributed by neighboring states, the bead usually must ``climb the hill'' of the current potential before the chemical dynamics ``catch up'' and switch to the next state. This forms an effective barrier region at these intersections. 

\begin{figure*}[th!]
    \centering
    \includegraphics[width=0.8\textwidth]{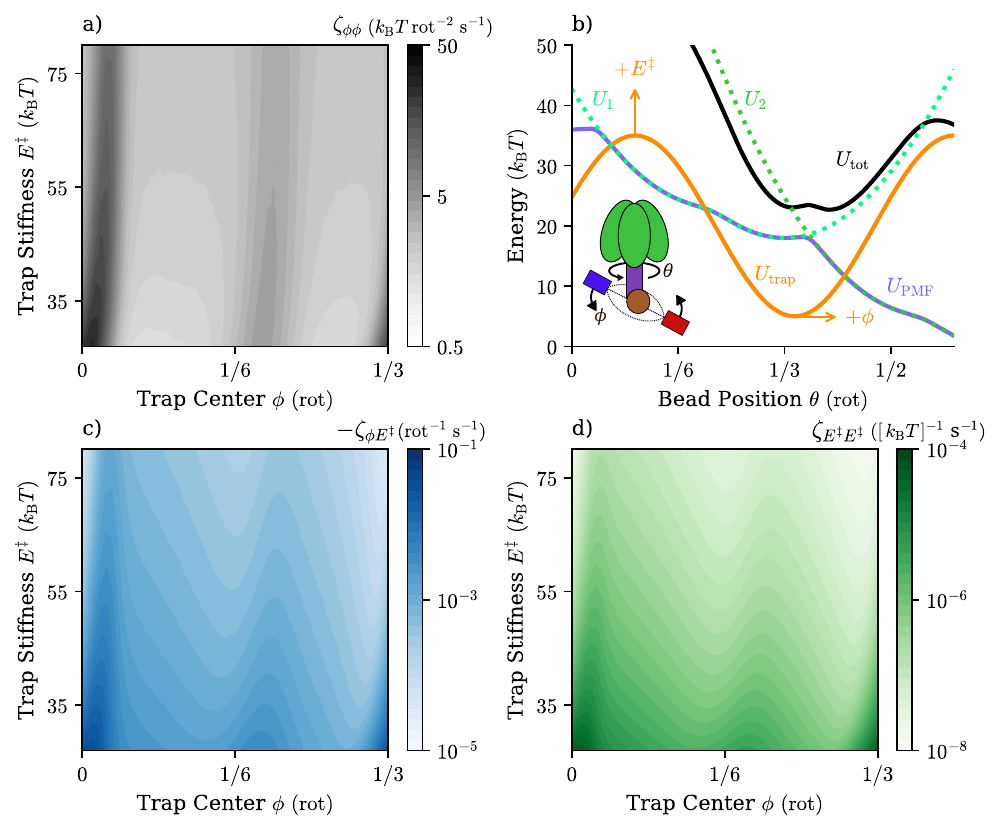}
    \caption{
    (a,c,d) Contour maps for the entries of the friction matrix as a function of the magnetic trap's controllable parameters (trap center $\phi$ and trap stiffness $\Eddag$) for the fast-chemistry limit ($\Gamma \to \infty$). (b) Schematic of the different contributions to the bead potential. Green dashed curves show potentials $U_1$ and $U_2$ for two neighboring chemical states. Blue solid curve: potential of mean force $U_{\textrm{PMF}}$ for fast chemistry. Solid orange curve: magnetic trap with $\phi = 0.35$~rot and $\Eddag = 30~k_{\rm B} T$. Solid black curve: total bead potential $U_{\mathrm{tot}} = U_{\mathrm{trap}} + U_{\mathrm{PMF}}$. Inset: schematic of the experiment.}
    \label{fig:PMF-friction}
\end{figure*}

In the high-concentration (fast-chemistry) case, $\Gamma \to \infty$, so mechanical rotation is slow compared to exchange between different potential landscapes due to chemical dynamics. The bead effectively experiences a single potential of mean force resulting from coarse-graining all rotational states~\cite{kawaguchiNonequilibriumDissipationfreeTransport2014},
\begin{equation}
    \beta U_{\mathrm{PMF}}(\theta) = -\ln \sum_{n=-\infty}^{\infty} e^{-\beta[U_n(\theta) - n\Delta \mu_{\mathrm{ATP}}]}.
\end{equation}
The barrier region appears explicitly in this case. Figure~\ref{fig:PMF-friction}(b) shows a schematic of the potentials.

The magnetic trap imposes on the bead the sinusoidal potential
\begin{equation}
    \beta U_\mathrm{trap}(\theta \,|\, \Eddag, \phi) = -\tfrac{1}{2}\Eddag \cos{2(\theta-\phi)},
\end{equation}
for trap stiffness $\Eddag$ and trap center $\phi$, and hence with minima at $\theta=\phi\pm m\pi$ for integer $m$~\cite{toyabeNonequilibriumEnergeticsSingle2010,toyabeThermodynamicEfficiencyMechanochemical2011,toyabeSingleMoleculeThermodynamics2015,nakayamaOptimalRectificationForwardCurrent2021,nakayamaAsymmetricEnzymeKinetics2025,guptaOptimalControlF1ATPase2022}. 

The bead evolves according to \add{the overdamped Langevin equation}
\begin{equation}
    \dot{\theta} = -\beta D\,\partial_\theta[U_n(\theta) + U_\mathrm{trap}(\theta \,|\, \Eddag, \phi)] + \sqrt{2D}\,\eta(t), \label{eq:evolution}
\end{equation}
with diffusion constant $D = 0.347$ $\mathrm{rot^2/s}$ and standard Gaussian white noise $\eta(t)$ with zero mean and unit variance. When $\Gamma \to \infty$, $U_n(\theta)$ is replaced with $U_\mathrm{PMF}(\theta)$.

The optimal control problem is to seek a \emph{control protocol}, a schedule for manipulating the control parameters $\phi$ and $\Eddag$, that minimizes the excess work done in moving the trap center $1/3$ of a rotation. In the near-equilibrium limit, this excess work is 
\begin{equation}
    \ev{W_{\textrm{ex}}^{\rm LR}} = \int_0^{t_{\textrm{prot}}} \frac{\partial \vec{\lambda}}{\mathrm{d}t}^T \zeta(\vec{\lambda}) \, \frac{\partial \vec{\lambda}}{\mathrm{d}t} \, \mathrm{d}t, \label{eq:excess_work}
\end{equation} 
for control-parameter vector $\vec{\lambda}$, protocol duration $t_{\textrm{prot}}$, and generalized friction matrix $\zeta$ defined as the time integral of the averaged covariance matrix of the conjugate forces:
\begin{equation}
    \zeta_{j\ell}(\vec{\lambda}) = \beta\int_0^\infty \ev{\delta \tau_j(0) \delta \tau_\ell(t)}_{\vec{\lambda}} \mathrm{d}t. \label{eq:friction_cov}
\end{equation}
Here $\ev{\cdot}_{\vec{\lambda}}$ is the equilibrium average at static control parameters $\vec{\lambda}$, and $\delta \tau_j(t) = \tau_j(t) - \ev{\tau_j}_{\vec{\lambda}}$ is the deviation of the instantaneous conjugate force
\begin{equation}
    \tau_j \equiv -\partial_{\lambda_j} U_\mathrm{trap}(\theta \,|\, \Eddag, \phi),
\end{equation}
from its equilibrium average $\ev{\tau_j}_{\vec{\lambda}}$. The friction matrix can also be written as 
\begin{equation}
    \zeta_{j\ell}(\vec{\lambda}) = \beta \ev{\delta \tau_j \delta \tau_\ell}_{\vec{\lambda}} \, t^{\rm relax}_{j\ell}(\vec{ \lambda}), \label{eq:friction_relax}
\end{equation}
a product of the force (co-)variance $\ev{\delta \tau_j \delta \tau_\ell}_{\vec{\lambda}}$ and relaxation time
\begin{equation}
    t^{\rm relax}_{j\ell}(\vec{ \lambda}) \equiv \int_0^\infty \mathrm{d}t \frac{\ev{\delta \tau_j(0) \delta \tau_\ell(t)}_{\vec{\lambda}}}{\ev{\delta \tau_j \delta \tau_\ell}_{\vec{\lambda}}}.
\end{equation}
\add{In the context of this work, equilibrium refers to mechanical equilibrium of the bead position given a fixed set of control parameters (in contrast to the chemical equilibrium of the bath).}

\emph{Designed} protocols that minimize excess work in the near-equilibrium limit satisfy an Euler-Lagrange equation involving the friction matrix and have constant excess power [the integrand of Eq.~\eqref{eq:excess_work}]. Therefore, the problem of calculating optimal protocols using this framework is essentially a problem of calculating this friction matrix, which acts as a metric for paths through control-parameter space~\cite{sivakThermodynamicMetricsOptimal2012, blaberOptimalControlStochastic2023}.

Previous investigations of optimal control in stochastic thermodynamics have usually sought the optimal path between two fixed endpoints~\cite{aurellOptimalProtocolsOptimal2011,blaberOptimalControlStochastic2023}. However, rotary molecular motors like $\mathrm{F_1}$-ATPase have an energy landscape that is periodic in its rotational coordinate, so we instead seek the most efficient path across this rotational coordinate without fixing any control-parameter points.

\section{Methods} \label{sec:methods}
We seek the protocol that most efficiently translates the trap center $\phi$ by $1/3$ rot under the periodic constraint that $\Eddag(\phi) = \Eddag(\phi + 1/3~\mathrm{rot})$. Furthermore, we impose limits on the trap stiffness, $28~k_{\rm B} T\leq\Eddag\leq79~k_{\rm B} T$. Below the lower limit, equilibrium trajectories frequently escape the magnetic trap; the upper limit is reasonably attainable in experiment~\add{\cite{mishimaEfficientlyDrivingF$_1$2025}}. The optimization problem we seek to solve is therefore directly analogous to finding the shortest path (according to metric $\zeta$) across the periodic coordinate $\phi \in [0, 1/3)$ rot, while remaining within $\Eddag \in [28~k_{\rm B} T,79~k_{\rm B} T]$.

To compare the relative stiffnesses of the trap and $\mathrm{F_1}$'s rotational substates~\eqref{eq:state_coarsegrained}, the second-order Taylor expansion of the sinusoidal trap near $\theta = \phi$ is 
\begin{equation}
    \beta  U_{\rm{harm}}(\theta \,|\, \Eddag, \phi) = \frac{1}{2}\left[- \Eddag + k_h (\theta - \phi)^2\right],
\end{equation}
a potential with stiffness $k_h \equiv 8\pi^2 \Eddag/\mathrm{rot^2}$. We explore $\Eddag$ corresponding to $k_h \in [2210~k_{\rm B}T/\mathrm{rot}^2,~6240~k_{\rm B}T/\mathrm{rot}^2]$, i.e., a harmonic trap that is $\approx$3-8$\times$ stiffer than $\mathrm{F_1}$'s substates. \add{The upper limit matches experimental and simulation estimates of the rotor-stalk stiffness~\cite{sielaffDomainComplianceElastic2008,okunoRotationStructureFoF1ATP2011a,czubTorsionalElasticityEnergetics2011,okazakiElasticityFrictionPathway2015}, which, as one in a series of mechanical elements linking proton translocation to ATP synthesis, constitutes an upper bound on the total effective stiffness of the machine.}.

We calculate the friction matrix $\zeta$ at a discrete grid \add{of control-parameter values} on this domain by computing the conjugate torque deviation [Eq.~\eqref{eq:friction_cov}] of individual trajectories, simulated following Eq.~\eqref{eq:evolution} \add{with the control parameters held constant}~\cite{sivakEquilibriumMeasurementsNonequilibrium2012,louwerseMultidimensionalMinimumworkControl2022,guptaOptimalControlF1ATPase2022}. We throw out any trajectories that escape the local well ($\theta\in[\phi-\pi/2,\phi+\pi/2]$) induced by the magnetic trap. We average the estimated friction matrix over $\approx$100 surviving trajectories.

We use the string method~\cite{maraglianoStringMethodCollective2006, maraglianoOntheflyStringMethod2007, louwerseMultidimensionalMinimumworkControl2022, blaberEfficientTwodimensionalControl2022} to improve a control protocol, iteratively reducing the excess work predicted in the linear-response framework. We initialize with a naive protocol (constant $\mathrm{d}\phi/\mathrm{d}t$ and $\Eddag$) with $N=201$ points along this path, representing control-parameter values at fixed times separated by a constant interval $\delta t$. Roughly speaking, at each iteration of the string method, the control parameters at each time-point are ``nudged'' in a direction that reduces the excess work, by a combination of the gradient of the friction coefficient and a ``tension force'' from the neighboring points. In general, the string points do not lie exactly on control-parameter values where the friction matrix has been calculated, so we linearly interpolate the friction-matrix grid to approximate the friction at arbitrary points.

We seek periodically repeating protocols, so we use a periodic string method to calculate these optimal paths, treating the points at the ``ends'' of the protocol as being connected across the periodic boundary conditions~\cite{maraglianoOntheflyStringMethod2007}. We also ensure that the protocol remains within the calculated range of $\Eddag$ values by artificially inserting a friction-coefficient ``wall'' immediately below $\Eddag = 28~k_{\rm B} T$ and above $\Eddag = 79~k_{\rm B} T$. These walls are a copy of the friction coefficients at the calculated boundaries, but scaled by a hand-tuned prefactor to produce a local zero in the gradient of the friction matrix. The friction matrix therefore exerts no force on the string points at the boundary, so string points remain there unless the neighboring points contribute a net inward force. 

For cases with $\Gamma > 10~\mathrm{s^{-1}}$, we terminate the string method when the excess work is no longer significantly reduced by applying further iterations. For $\Gamma = 10~\mathrm{s^{-1}}$ (slow chemistry), the string method becomes stiff and eventually fails. In this case, the friction tensor varies rapidly, with a particularly strong gradient in the barrier region. The algorithm initially reduces the excess work by almost $3\times$ compared to the naive (constant-$\Eddag$) initial protocols. However, owing to instability in subsequent iterations, the string method fails to further improve the protocol. We therefore use the protocol that empirically minimizes the excess work across all iterations of the method.

Finally, each path through control-parameter space given by the string method is optimized by moving the string points along the given path to equalize the excess work for each discrete time interval $\delta t$. For the protocols considered here, this does not significantly change the total predicted excess work. We label the protocols that result from the combined string-method and along-the-path optimizations as ``center-stiffness'' designed protocols, since they dynamically vary both control parameters.

We compare the center-stiffness protocols to ``center-only'' designed protocols that have fixed $\Eddag$. Center-only protocols using $\Eddag = 30~k_{\rm B} T$ and $\Eddag = 60~k_{\rm B} T$ were analyzed in Ref.~\cite{guptaOptimalControlF1ATPase2022}. We calculate protocols of this type by applying the along-the-path optimization method to naive protocols, which in the continuous limit is equivalent to enforcing the proportionality $\mathrm{d}\phi/\mathrm{d}t \propto \zeta_{\phi\phi}^{-1/2}$.

\begin{figure*}[ht!]
    \centering
    \includegraphics[width=0.99\textwidth]{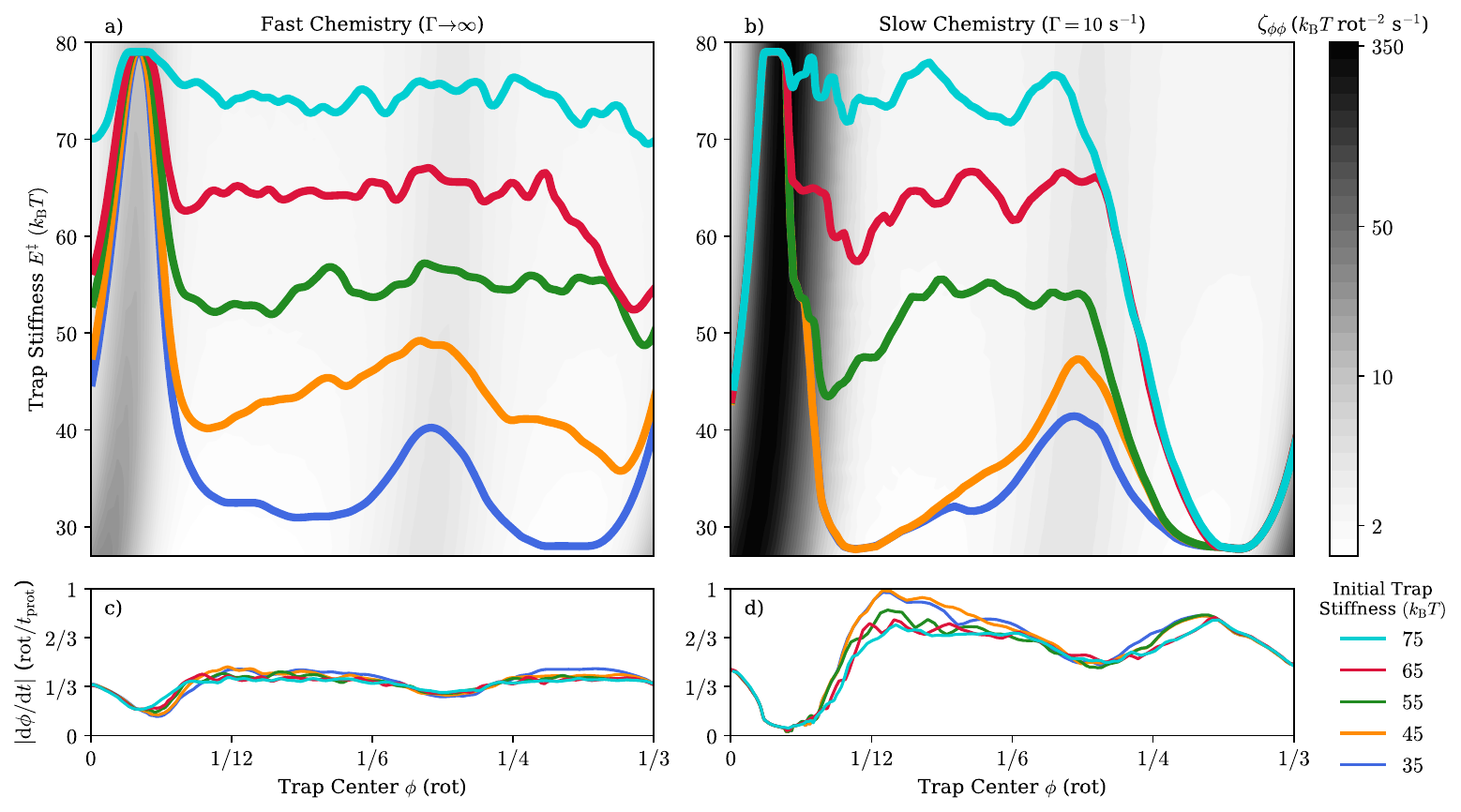}
    \caption{multiparameter designed protocols for fast (a,c) and slow (b,d) chemistry. The protocols were computed using the string method, initialized with naive protocols at initial trap stiffnesses $\Eddag$ specified in the legend. (a,b) The designed path through control-parameter space, overlaid on $\zeta_{\phi\phi}$. (c,d) Parametric plots of the trap speed $|\mathrm{d}\phi/\mathrm{d}t|$ and trap center $\phi$, illustrating that $\phi$ slows down where the friction coefficient is high.}
    \label{fig:combined-starting-heights}
\end{figure*}

We simulate experiments following naive and designed protocols with different protocol durations $t_{\textrm{prot}}$, and compare the average excess work to that predicted by the linear-response framework. 
The discretization of Eq.~\eqref{eq:evolution} is
\begin{equation}
    \theta_{i+1} = \theta_{i} - \beta D \, \partial_\theta U_{\rm tot}(\theta_{i}\,|\,\phi_i, \Eddag_i) \, \Delta t + \sqrt{2D\Delta t} \, \eta_{i}, \label{eq:discrete}
\end{equation}
where $U_{\rm tot} \equiv U_{\mathrm{trap}} + U_n$ (for fast chemistry, replace $U_n$ with $U_{\mathrm{PMF}}$) and $\eta_i$ is discrete Gaussian white noise with zero mean and unit variance. 
\add{When the switching potential $U_n$ associated with a discrete chemical state is active, at each timestep the identity of the current potential switches in the hydrolysis direction, $U_n \to U_{n+1}$ (synthesis direction, $U_n \to U_{n-1}$) with probability $R_n^+ \Delta t$ ($R_n^- \Delta t$), and remains unchanged with probability $1-(R_n^+ + R_n^-) \Delta t$.}

We evolve Eq.~\eqref{eq:discrete} with $\Delta t = 10^{-6}$ s and with $\phi_i$ and $\Eddag_i$ linearly interpolated between the calculated friction-matrix grid points. Each trajectory is initialized at $\theta_0$ drawn from the equilibrium distribution for $U(\theta|\phi_0,\Eddag_0)$. To remove the effect of the initial condition and measure the excess work in steady-state operation, the protocol is repeated $15$ times without interruption. Only the last full rotation (three protocol repetitions) contributes to the estimate of the excess work,
\begin{equation}
    W_{\mathrm{ex}} = \sum_{i=0}^{t_{\mathrm{prot}}/\Delta t - 1} W_i \pm \Delta F, \label{eq:simwork}
\end{equation}
for work 
\begin{equation}
    W_i = U_{\mathrm{trap}}\left(\theta_i\,|\, \phi_{i+1}, \Eddag_{i+1}\right) - U_{\mathrm{trap}}\left(\theta_i\,|\, \phi_{i}, \Eddag_{i}\right)
\end{equation}
during a single discrete timestep. $\Delta F = 3\Delta \mu_{\textrm{ATP}} = 54~k_{\rm B}T$ is the total equilibrium free-energy change during synthesis, and the $\pm$ signs refer to ATP hydrolysis and synthesis, respectively. We average over $10^5$ trajectories for each protocol duration, in both \stkout{hydrolysis}\add{synthesis} ($\phi$ \stkout{increasing}\add{decreasing}) and \stkout{synthesis}\add{hydrolysis} directions ($\phi$ \stkout{decreasing}\add{increasing}) \add{to respectively mimic the natural operation of either F-ATPases or the ATP-hydrolytic, ion-pumping vacuolar (V-) ATPases~\cite{collinsRegulationFunctionVATPases2020}}. Any simulation runs in which the bead escapes the local trap are ignored in calculating the average excess work, but we only consider durations for which the bead remains within the trap in more than 80\% of trajectories \add{(in the vast majority of conditions, less than a few percent escape)}.

\section{Results} \label{sec:results}

\begin{figure}[t]
\centering
\includegraphics[width=\columnwidth]{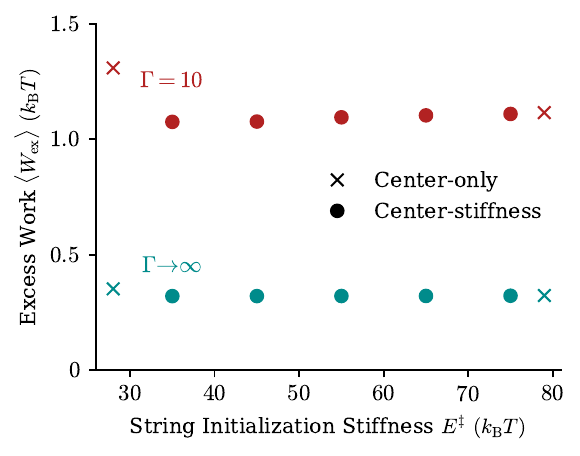}
\caption{Linear-response prediction~\eqref{eq:excess_work} for excess work $\ev{W_{\mathrm{ex}}}$ during a one-second protocol. Crosses: center-only designed protocols with fixed stiffness $\Eddag=28~k_\mathrm{B}T$ and $\Eddag=79~k_{\mathrm{B}}T$. Circles: center-stiffness protocols from Fig.~\ref{fig:combined-starting-heights} (calculated with the string method initialized at trap stiffnesses specified in the legend). Red, blue: slow chemistry ($\Gamma=10$) and fast chemistry ($\Gamma\to\infty$), respectively.}
\label{fig:heights_vs_works}
\end{figure}

\begin{figure*}[ht!] 
    \centering 
    \includegraphics[width=\textwidth]{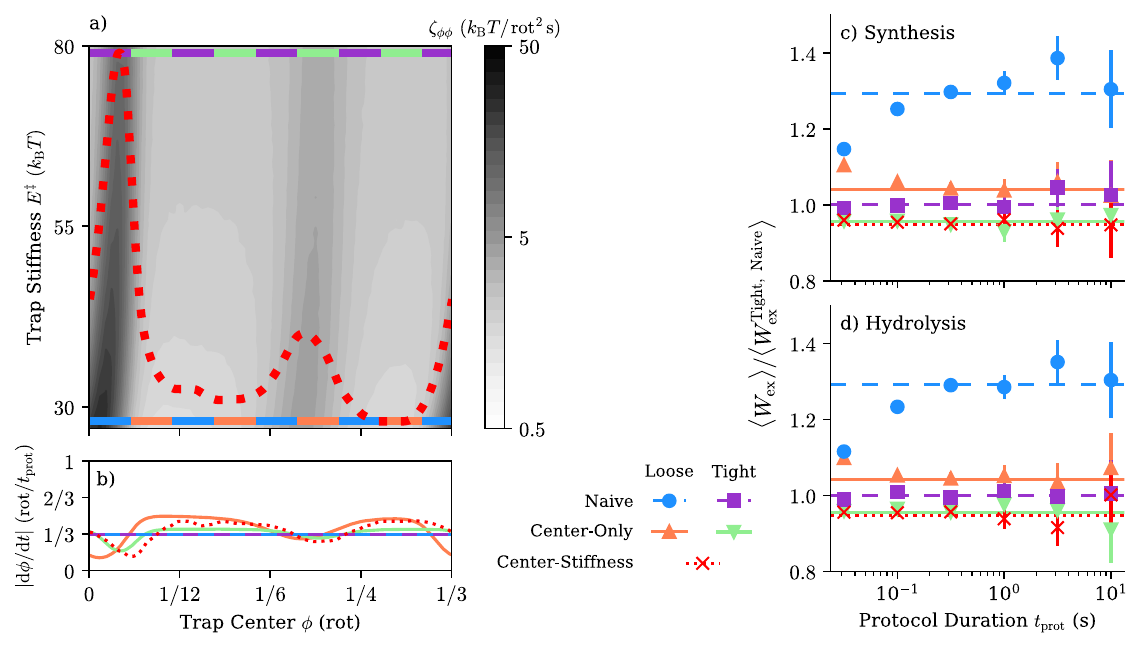}
    \caption{Excess work $\ev{W_\textrm{ex}}$ in naive, single-parameter designed, and multiparameter designed protocols, for fast chemistry ($\Gamma\to\infty$). (a) Protocol path through control-parameter space, overlaid on $\zeta_{\phi\phi}$ (grayscale). (b) Trap speed $|\mathrm{d}\phi/\mathrm{d}t|$ as a function of $\phi$. (c,d) Excess work $\ev{W_\textrm{ex}}$ predicted by linear response~\eqref{eq:excess_work} (lines) and calculated by Langevin simulations (points). For clarity, all the data have been normalized to the linear-response excess work predicted by the tight-trap naive protocol $\ev{W_{\textrm{ex}}^{\textrm{Tight,~Naive}}}$ at the corresponding $t_{\mathrm{prot}}$ (purple line). ``Loose'' and ``tight'' naive protocols are shown with \stkout{solid} \add{dashed} blue and purple lines (circles and squares), respectively. Their single-parameter designed counterparts are shown with \stkout{dashed} \add{solid} orange and green curves (upward- and downward-pointing triangles), respectively, while a multiparameter designed protocol is represented by a dotted red curve (crosses).}  
    \label{fig:protocol-works}
\end{figure*}

Figure~\ref{fig:PMF-friction} shows contour maps of the elements of the friction matrix for the case of fast chemistry ($\Gamma \to \infty$). The (trap center)-(trap center) component $\zeta_{\phi\phi}$ [Fig.~\ref{fig:PMF-friction}(a)] at a given trap stiffness is equivalent to that calculated in the single-parameter case for a set trap stiffness, quantifying the excess work dissipated in (slowly) changing only the trap-center component~\cite{guptaOptimalControlF1ATPase2022}. This component provides all the information needed to compute a center-only designed protocol. The other diagonal element is the (trap stiffness)-(trap stiffness) component $\zeta_{\Eddag\Eddag}$ [Fig.~\ref{fig:PMF-friction}(d)], which quantifies the excess work dissipated when (slowly) tightening or loosening the magnetic trap. Finally, the off-diagonal component $\zeta_{\Eddag\phi}$ [Fig.~\ref{fig:PMF-friction}(c)] quantifies any extra or reduced cost associated with (slowly) simultaneously changing the trap center and the trap stiffness. Notice that $\zeta_{\Eddag\phi} < 0$, meaning that this component reduces (compared to that suggested by the diagonal components alone) the excess work when $\mathrm{d}\phi/\mathrm{d}t$ and $\mathrm{d}\Eddag/\mathrm{d}t$ have the same sign.

Like $\zeta_{\phi\phi}$ in Ref.~\cite{guptaOptimalControlF1ATPase2022}, all three friction coefficients are greatest in magnitude when the trap-center component $\phi$ is in the barrier region, where the potentials from neighboring chemical states are comparable. In this region, both the force variance and force relaxation time [Eq.~\eqref{eq:friction_relax}] are largest due to the presence of two local minima in the energy landscape that the bead experiences~\cite{guptaOptimalControlF1ATPase2022}.

Figures~\ref{fig:combined-starting-heights}(a) and \ref{fig:combined-starting-heights}(b) show contour plots of $\zeta_{\phi\phi}$ (perhaps the most immediately physically intuitive component) in the respective cases of $\Gamma$$\to$$\infty$ (fast chemistry) and $\Gamma=10~\mathrm{s^{-1}}$ (slow chemistry). \add{These two conditions are comparable~\cite{guptaOptimalControlF1ATPase2022} to the fastest and slowest chemical dynamics experimentally observed in Ref.~\cite{toyabeThermodynamicEfficiencyMechanochemical2011}}. The two cases have a similar structure, but $\zeta_{\phi\phi}$ is much larger near the switching barrier for $\Gamma=10~\mathrm{s^{-1}}$. Physically, this slow-chemistry limit means that the bead tends to be caught on one side of the barrier for longer times, increasing the relaxation time~\cite{guptaOptimalControlF1ATPase2022}. Overlaid on these contour plots are center-stiffness designed protocols calculated using the string method from Sec.~\ref{sec:methods}, initialized with naive protocols (constant $\mathrm{d}\phi/\mathrm{d}t$ and $\Eddag$) at different trap stiffnesses equally spaced between $35~k_{\rm B} T$ (blue) and $75~k_{\rm B} T$ (cyan); the displayed final protocols visually preserve the stiffness-ordering of their initializations. Figures~\ref{fig:combined-starting-heights}(c) and \ref{fig:combined-starting-heights}(d) are parametric plots showing the trap-center velocity $\mathrm{d}\phi/\mathrm{d}t$ vs.\ trap center.

The center-stiffness protocols all start from different initializations but should theoretically converge to an ``optimal'' protocol on their friction-matrix landscape. For fast chemistry, they predict excess works differing by less than $0.6\%$, while for slow chemistry they differ by a maximum of $3\%$ (Fig.~\ref{fig:heights_vs_works}). Here, we cannot quantify how near or far our final designed protocols are to that optimum; however, it is notable that we have initialized our optimization procedure from different protocols with very different initial excess works and update them until instabilities prevent further optimization, and still find very similar values for excess work at the end of the procedure. The features they share are therefore likely important for minimizing the excess work (and conversely, aspects that vary are likely unimportant).

All the designed protocols in Fig.~\ref{fig:combined-starting-heights} slow the trap center (reduce $\mathrm{d}\phi/\mathrm{d}t$) and tighten the trap (increase $\Eddag$) near the barrier; for $\Gamma = 10~\mathrm{s^{-1}}$, the trap is also maximally loose immediately to one side of the barrier (between $\phi$$\approx$1/4 and 1/3 rot). The protocols slow down more when crossing the barrier under slow chemistry than under fast chemistry, due to the greater variation of $\zeta_{\phi\phi}$.

Strong agreement between these designed protocols in the barrier region---and significant variation away from it---can be understood by comparing the absolute scale of the (trap-center)-(trap-center) friction coefficient $\zeta_{\phi\phi}$ across the range shown in Fig.~\ref{fig:PMF-friction}(a). The peak value of $\zeta_{\phi\phi}$ at fixed maximal $\Eddag=79~k_{\rm B}T$ differs from that at fixed minimal $\Eddag=28~k_{\rm B}T$ by about $15~k_{\rm B} T/\rm{rot^2\,s}$. Away from the barrier region, the absolute scale of $\zeta_{\phi \phi}$ falls well below this, so its variation across either control parameter leaves little for a protocol to exploit. In other words, small variations of the chosen protocol in the barrier region come at a much greater cost than variations away from the barrier.

Friction coefficients for intermediate $\Gamma$ ($\Gamma = 100~\mathrm{s^{-1}}$ and $1000~\mathrm{s^{-1}})$ were also calculated and differ mostly in the height of $\zeta_{\phi\phi}$ at the barrier (interpolating between the fast- and slow-chemistry cases); see Appendix~\ref{app:intermediate}. Designed protocols in those cases look similar, with the trap center at the barrier moving slower than for fast chemistry, but faster than for slow chemistry.

To quantify the impact that dynamic control of the trap stiffness has on reducing the excess work, we compare the most efficient center-stiffness protocol [blue curve in Fig.~\ref{fig:combined-starting-heights}(a)] to naive and center-only designed protocols (which use only dynamic control over the trap center, keeping $\Eddag$ fixed) constrained to either the top ($\Eddag = 79~k_{\rm B}T$) or bottom ($\Eddag = 28~k_{\rm B} T$) of the range of permitted trap stiffness. Here, we discuss only the case of fast chemistry, but the qualitative results are preserved in cases using slower chemistry [Fig.~\ref{fig:combined-starting-heights}(b) and Appendix~\ref{app:intermediate}].

Figure~4(a) shows the path each protocol takes through control-parameter space, overlaid on $\zeta_{\phi\phi}$. Figure~4(b) displays the trap-center velocity as a function of trap center. Figures~4(c) and 4(d) display $\ev{W_{\mathrm{ex}}}$ as a function of protocol duration during synthesis (negative $\mathrm{d}\phi/\mathrm{d}t$) and hydrolysis (positive $\mathrm{d}\phi/\mathrm{d}t$)\add{, normalized to the predicted linear-response excess work $\ev{W_{\textrm{ex}}^{\textrm{Tight,~Naive}}}$ for the tight-trap naive protocol}. \add{We present the results in this normalized form to emphasize the relative performance between the protocols across orders of magnitude in protocol duration. The predicted linear-response $\ev{W_{\mathrm{ex}}}$ scales like $1/t$ and therefore diverges at short protocol times; however, the true $\ev{W_{\mathrm{ex}}}$ asymptotes to the (constant, protocol-independent) value for the case of an instantaneous protocol, given by the Kullback-Leibler divergence~\cite{largeOptimalDiscreteControl2019,blaberStepsMinimizeDissipation2021} Hence, in Figs.~4(c,d), linear-response predictions (normalized to the tight-trap naive prediction) appear as horizontal lines, and at asymptotically short protocol durations the (normalized) true excess work decreases.}

Comparing the linear-response approximations, the ``loose naive'' protocol performs the worst (as expected), while its center-only counterpart reduces excess work by $\approx$1.24$\times$. Interestingly, both of these loose protocols perform worse than the tight naive protocol. The tight center-only and the center-stiffness protocols perform similarly, both reducing the excess work by $\approx$1.05$\times$ from the tight naive protocol. Figure~\ref{fig:heights_vs_works} shows that the tight center-only protocol predicts essentially the same excess work as all of the center-stiffness protocols with different initializations in Fig.~\ref{fig:combined-starting-heights}(a), so the additional degree of dynamic control exercised by the multiparameter protocol does little to reduce the predicted excess work. In the slow-chemistry case, the extra degree of control reduces the excess work by $\approx$1.03$\times$ (see Appendix~\ref{app:slowsims}); the additional utility here may be due to the larger variation in the components of the friction matrix.

For most of the protocol types and durations considered in the fast-chemistry case, the excess work measured in the simulations is comparable to the linear-response approximations for both the synthesis and hydrolysis directions. Compared to the linear-response prediction, the relative performance of the loose naive protocol appears to improve for very short protocol durations, but this is probably because it has begun to approach the excess work in a near-instantaneous protocol (i.e., the maximum excess work possible in a single step; cf.\ Fig.~6 in Ref.~\cite{blaberOptimalControlStochastic2023}).

\section{Discussion}

In this work, we designed energy-efficient multiparameter control protocols in the near-equilibrium regime for a model of $\mathrm{F_1}$-ATPase -- one of biology's most ubiquitous rotary motors. We found that the intuition gained from previous works about how to design an efficient protocol (control potentials should be tight and slow down near energy barriers) also holds here~\cite{sivakThermodynamicGeometryMinimumdissipation2016,luceroOptimalControlRotary2019,blaberEfficientTwodimensionalControl2022,guptaOptimalControlF1ATPase2022,loosUniversalSymmetryOptimal2024}. Furthermore, we found that protocols with a static tight trap and dynamic control of only the trap center can satisfy the key aspects of efficiency and perform as well as their multiparameter counterparts. The barrier region (where $\zeta_{\phi\phi}$ is highest) provides the greatest opportunity for saving excess work, which efficient protocols of all types exploit in essentially the same way.

Near the barrier, the trap must be tight, so the trap stiffness here is a ``stiff'' mode~\cite{machtaParameterSpaceCompression2013}. Even for naive protocols, choosing a trap stiffness at the top of our considered range (instead of the bottom) satisfies this requirement and results in a large reduction in the excess work. However, the trap stiffness away from the barrier is a ``sloppy'' mode; its dynamic optimization can only achieve comparatively small work savings. In this case, then, the simplest strategy to achieve efficient driving is to make the trap as tight as possible (reducing $\zeta_{\phi\phi}$ as much as possible) and follow a center-only designed protocol, essentially surrendering dynamic control of the second coordinate. 

In contrast to the relatively small efficiency gain found here for multiparameter protocols over single-parameter protocols, Ref.~\cite{blaberEfficientTwodimensionalControl2022} showed for a double-well potential that the second dimension of control (directly analogous to $\Eddag$) could make protocols $3.5\times$ more efficient compared to single-parameter optimized counterparts with the same endpoints. This is a different optimization problem from the one we solve here -- they sought the optimal path between two points in control-parameter space, whereas we seek the optimal path around a cylindrical control-parameter space without requiring the protocol to pass through any specified point. In particular, they only compared a single-parameter protocol analogous to our ``loose center-only'' protocol to their multiparameter designed protocol. Additionally, the control protocols considered in their model never reached their imposed boundaries on trap stiffness. Here, by contrast, including the second dimension of control reduces the predicted $\ev{W_{\mathrm{ex}}}$ by only $\approx$1.1$\times$ compared to the loose center-only protocol. It is likely that more savings could be achieved in this model if the (experimentally motivated) upper bound on the trap stiffness was increased. Furthermore, the barrier created in this model is smaller than that used in Ref.~\cite{blaberEfficientTwodimensionalControl2022}, which reduces opportunities for savings through dynamic control~\cite{sivakThermodynamicGeometryMinimumdissipation2016}.

In the broader context of optimal control of stochastic systems, this small efficiency gain is a reminder that not all control parameters are created equally -- especially under physical constraints. Adding an additional way to dynamically adjust a stochastic system's distribution is not guaranteed to result in a sharp gain in efficiency. Based on our results, a reasonable design strategy for biological ATP synthase is to exercise tight control over its rotor stalk throughout its entire operating cycle.

\add{In accordance with physical intuition, these results suggest that more rigid coupling may prove beneficial in connecting (biological or synthetic) nanomachine subunits;} \add{however, the}\stkout{The} design of an organism's molecular machines yields a selective advantage only if it results in an efficiency gain over their entire lifecycle -- including both construction and operation. When the operational efficiency gain from adding a new degree of control is small, these advantages may be outweighed by the costs of building the presumably more complex mechanism for this control. Even evolutionarily finding such a mechanism may be challenging, since ATP synthase's essential role may make it too tightly constrained to allow for drastic innovations~\cite{kucharczykMitochondrialATPSynthase2009,cilibertiInnovationRobustnessComplex2007}.

In this near-equilibrium framework, the periodicity of the control problem boils down to ensuring continuity of the control parameters between cycles. Biological molecular machines like $\mathrm{F_1}$, however, operate far from equilibrium under strong gradients. Away from the slow-driving regime~\cite{blaberStepsMinimizeDissipation2021} and in true steady-state operation, different control protocols from these linear-response results may be more efficient. Theory on how to address this steady-state, far-from-equilibrium regime would therefore be useful in further understanding efficient design principles of molecular machines.

Though the linear-response approach provides a tractable method, it is still computationally costly to go beyond one or two control parameters~\cite{SawchukDecompositions} when the potential landscape is complicated~\cite{luceroOptimalControlRotary2019}. There is a need for more theory on isolating (in an initial characterization before full analysis) which control parameters are useful to vary dynamically, which are important to set to a particular constant value, and which are relatively unimportant. \stkout{Another approach to this could be through machine-learning techniques, which have already proven fruitful in designing efficient protocols near and far from equilibrium~\cite{whitelamHowTrainYour2023,engelOptimalControlNonequilibrium2023}.}

\add{Machine-learning techniques represent a complementary strategy for investigating both far-from-equilibrium control and many-parameter control. This approach has already proven fruitful in designing efficient protocols near and far from equilibrium~\cite{whitelamHowTrainYour2023,engelOptimalControlNonequilibrium2023}.}

The control protocols discussed here are all examples of open-loop control, where the system output (i.e., the bead position) is not known or used. However, biological ATP synthase may use, for example, the tension in its rotor and/or stator stalks to track the difference between $\mathrm{F_o}$ and $\mathrm{F_1}$'s rotational positions~\cite{jungeTorqueGenerationElastic2009,guoStructureATPSynthase2022,sobtiChangesCentralStalk2023}. This would allow it to effectively apply a closed-loop (feedback) control protocol~\cite{bechhoeferFeedbackPhysicistsTutorial2005,sagawaNonequilibriumThermodynamicsFeedback2012a}. Regardless of which control strategy ATP synthase uses in reality, modeling efficient feedback control for it could yield broader insights into effective design principles for molecular machines.

\acknowledgments
The authors thank Jordan Sawchuk (SFU Physics), Shoichi Toyabe (Tohoku University Applied Physics), and Deepak Gupta (IIT Indore Physics) for helpful discussions. This work was supported by a Natural Sciences and Engineering Research Council of Canada (NSERC) CGS Master's Fellowship (W.C.W.), an NSERC Discovery Grant RGPIN-2020-04950 (D.A.S.), and a Tier-II Canada Research Chair CRC-2020-00098 (D.A.S.), and was enabled in part by support provided by the Digital Research Alliance of Canada (alliancecan.ca).

\appendix

\section{Designed Protocols with Intermediate Chemical Rates} \label{app:intermediate}

\begin{figure*}[ht!]
    \centering
    \includegraphics[width=0.9\textwidth]{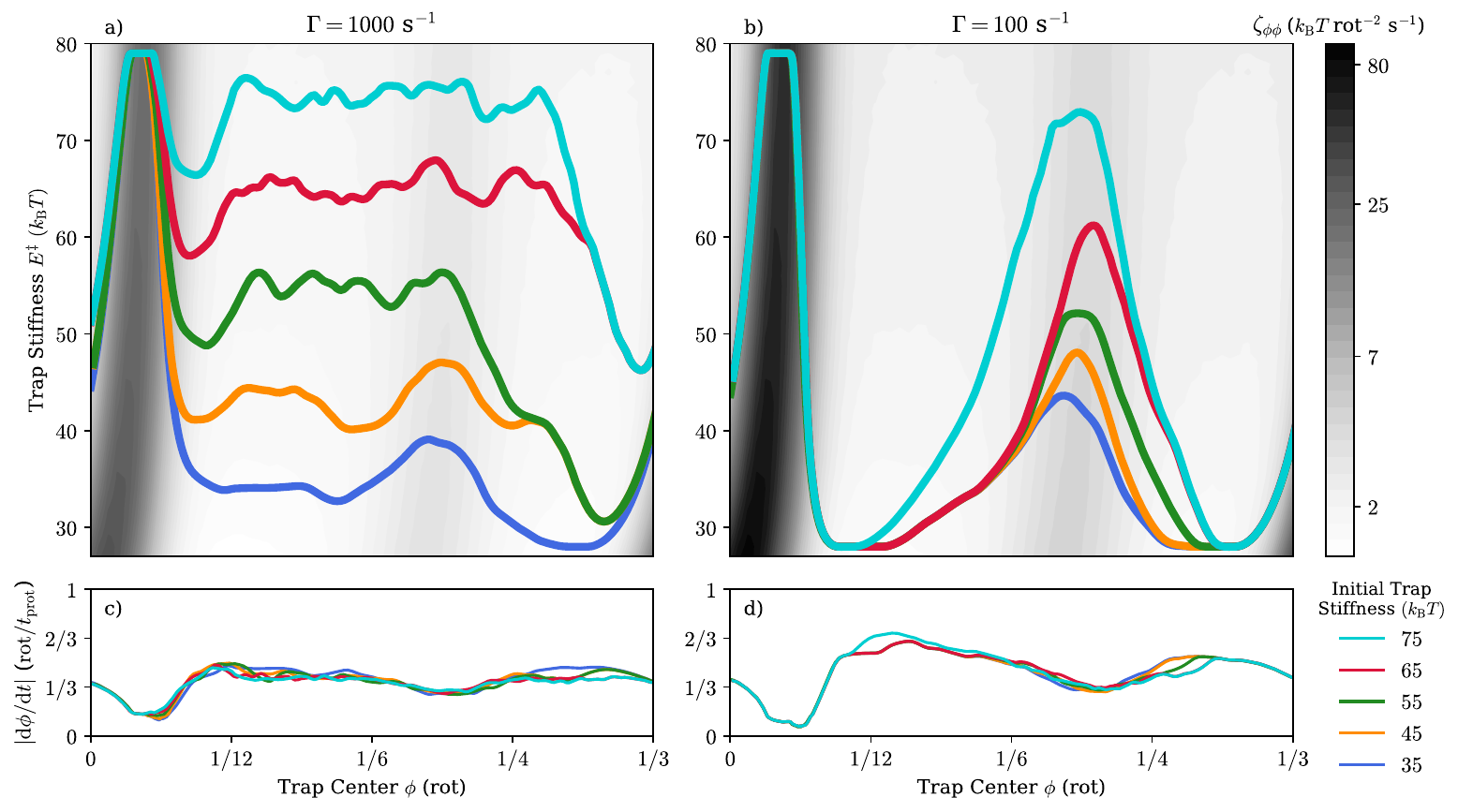}
    \caption{multiparameter designed protocols for intermediate values of the rate constant: $\Gamma = 100~\mathrm{s}^{-1}$ (a,c) and $\Gamma = 1000~\mathrm{s}^{-1}$ (b,d). (a,b) The designed path through control-parameter space, overlaid on $\zeta_{\phi\phi}$. (c,d) Parametric plots of the trap speed $|\mathrm{d}\phi/\mathrm{d}t|$ and trap center $\phi$.}
    \label{fig:mid-gamma-protocols}
\end{figure*}

Designed protocols were also calculated for intermediate values of the rate constant, $\Gamma = 100~\mathrm{s}^{-1}$ and $\Gamma = 1000~\mathrm{s}^{-1}$. Figure~$\ref{fig:mid-gamma-protocols}$ reproduces Fig.~\ref{fig:combined-starting-heights}, but with $\Gamma = 1000~\mathrm{s}^{-1}$ (left column) and $\Gamma = 100~\mathrm{s}^{-1}$  (right). In both cases, the protocols for a given $\Gamma$ predict excess works differing by less than $0.6\%$. Like the slow- and fast-chemistry cases in the main text, all protocols satisfy the stiff mode by tightening the trap near the barrier. Away from the barrier, the trap stiffness is less important in reducing the excess work.

%\begin{widetext}

\begin{figure*}[ht!]
    \centering
    \includegraphics[width=0.9\textwidth]{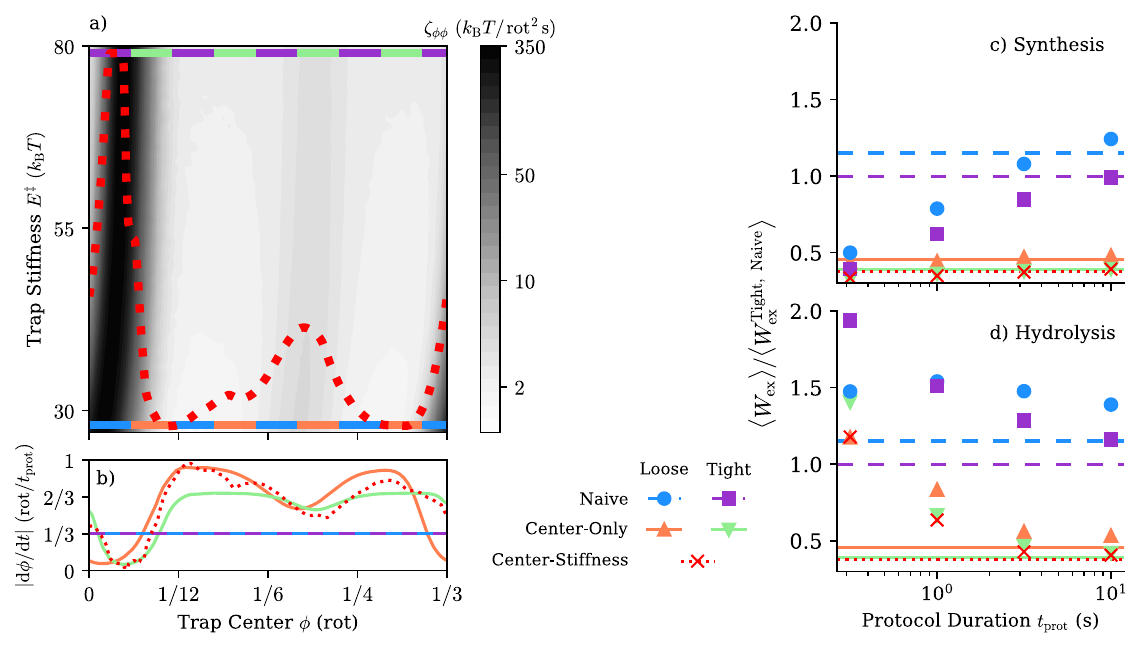}
    \caption{Excess work $\ev{W_\textrm{ex}}$ in naive, center-only, and center-stiffness designed protocols for the case of slow chemistry ($\Gamma = 10~\mathrm{s^{-1}}$). (a) Protocol path through control-parameter space, overlaid on $\zeta_{\phi\phi}$. (b) Trap speed $|\mathrm{d}\phi/\mathrm{d}t|$ as a function of $\phi$. (c,d) Excess work $\ev{W_\textrm{ex}}$ predicted by Eq.~\eqref{eq:excess_work} (lines) and calculated by Langevin simulations (points). For clarity, all the data have been normalized to the linear-response excess work predicted by the tight-trap naive protocol $\ev{W_{\textrm{ex}}^{\textrm{Tight,~Naive}}}$ at the corresponding $t_{\mathrm{prot}}$ (purple line). ``Loose'' and ``tight'' naive protocols are shown with \stkout{solid} \add{dashed} blue and purple lines (circles and squares), respectively. Their single-parameter designed counterparts are shown with \stkout{dashed} \add{solid} orange and green curves (upward- and downward-pointing triangles), respectively, while a multiparameter designed protocol is represented by a dotted red curve (crosses).  Error bars in (c,d) are smaller than symbols.}  
    \label{fig:gamma10-protocol-works}
\end{figure*}
%\end{widetext}

\clearpage
\section{Slow-Chemistry Simulations} \label{app:slowsims}

Figure \ref{fig:gamma10-protocol-works} compares in the slow-chemistry case ($\Gamma = 10~\mathrm{s^{-1}}$) the excess work predicted by linear response to the true values from simulation. The center-only designed protocol with fixed tight stiffness and the center-stiffness designed protocol reduce the predicted $W_\textrm{ex}$ compared to the tight-naive protocol by $\approx$2.57$\times$ and $2.66\times$, respectively. More savings are possible here compared to the fast-chemistry case because of the larger variation across the control-parameter space of the components of the friction matrix. Still, the additional degree of dynamic control exercised by the multiparameter-designed protocol only decreases the excess work by $1.03\times$. 

When driving ATP synthesis [Fig.~\ref{fig:gamma10-protocol-works}(c)], all the designed protocols for the longest durations produce $\ev{W_\textrm{ex}}$ similar to the linear-response predictions. For shorter durations, the naive protocols produce $\ev{W_\textrm{ex}}$ lower than the linear-response prediction because in the fast-driving limit the actual excess work asymptotes to the finite value of an instantaneous protocol~\cite{blaberEfficientTwodimensionalControl2022} rather than diverging as in the linear-response approximation.

When driving ATP hydrolysis [Fig.~\ref{fig:gamma10-protocol-works}(d)], all of the simulations yield larger $\ev{W_\textrm{ex}}$ than the linear-response predictions, consistent with the results presented in Fig.~S4 of Ref.~\cite{guptaOptimalControlF1ATPase2022}. Unlike the ATP synthesis rate, the hydrolysis rate in our model is constant and independent of the bead position, which means that the chemical dynamics are more likely to lag behind the trap center. The bead, therefore, experiences larger forces from the trap (i.e., the trap does more work). In this respect, the bead is further from equilibrium when operating in the (chemically driven) hydrolysis direction. We emphasize that this results from the model's asymmetric splitting of chemical rates, chosen for their good fit of experimental results for free hydrolysis (without mechanical driving)~\cite{kawaguchiNonequilibriumDissipationfreeTransport2014,guptaOptimalControlF1ATPase2022} -- hence the tight correspondence in the fast-chemistry case (Fig.~\ref{fig:protocol-works}), where the mechanical motion is insensitive to the details of the chemical switching rates.

\bibliography{Sivak}

\end{document}